\documentstyle[aas2pp4]{article}
\def\w{$w(\theta )\ $}
\def\puncspace{\ifmmode\,\else{\ifcat.\C{\if.\C\else\if,\C\else\if?\C\else%
\if:\C\else\if;\C\else\if-\C\else\if)\C\else\if/\C\else\if]\C\else\if'\C%
\else\space\fi\fi\fi\fi\fi\fi\fi\fi\fi\fi}%
\else\if\empty\C\else\if\space\C\else\space\fi\fi\fi}\fi}
\def\SP{\let\\=\empty\futurelet\C\puncspace}
\def\h1{$h^{-1}$\SP}

\def\lessapprox{\,\raise 0.6ex\hbox{$<$}\kern -0.75em\lower 0.47ex
    \hbox{$\sim$}\,}
\def\largapprox{\,\raise 0.6ex\hbox{$>$}\kern -0.75em\lower 0.47ex
    \hbox{$\sim$}\,}
\def\lsim{\lessapprox}
\def\gsim{\largapprox}
\def\ni{\noindent}

\def\chrger than approximately
\def\largapprox{\,\raise 0.6ex\hbox{$>$}\kern -0.75em\lower 0.47ex
    \hbox{$\sim$}\,}
\def\lsim{\lessapprox}
\def\gsim{\largapprox}
\def\ni{\noindent}

\def\be{\begin{equation}}
\def\ee{\end{equation}}
\def\bea{\begin{eqnarray}}
\def\eea{\end{eqnarray}}

\def\eps2{{\epsilon^2}}

\def\eg{{\sl eg.\ }}
\def\ie{{\sl i.e.\ }}
\def\etal{{\sl et al.\ }}
\def\b91{Blandford \etal (1991)}

\def\half{\frac{1}{2}}

\def\go{\mathrel{\raise.3ex\hbox{$>$}\mkern-14mu
             \lower0.6ex\hbox{$\sim$}}}
\def\lo{\mathrel{\raise.3ex\hbox{$<$}\mkern-14mu
             \lower0.6ex\hbox{$\sim$}}}

\begin{document}
\title{Clustering of Galaxies in the Hubble Deep Field}

\author{Jens V. Villumsen}
\affil{ Max-Planck-Institut f\"ur Astrophysik,
Karl-Schwarzschild-Str. 1, D--85740 Garching, Germany}

\author{Wolfram Freudling}
\affil{Space Telescope--European Coordinating Facility and European Southern
Observatory, Karl--Schwarzschild--Str. 2, D--85748 Garching b. M\"unchen,
Germany}

\author{Luiz N. da Costa}
\affil{European Southern Observatory, Karl--Schwarzschild--Str. 2, D--85748
Garching b. M\"unchen, Germany and Observat\'orio Nacional, Rua Jos\'e
Cristino 77, Rio de Janeiro,
Brazil}

%\maketitle

\begin{abstract} 

We compute the two-point angular correlation function \w for a sample
of $\sim$ 1700 galaxies to a magnitude-limit equivalent to
$R~\sim~29.5$ using a catalog derived from the Hubble Deep Field
images.  A non zero value of \w is measured down to $R=29.0$. The
amplitude of \w at the bright magnitude limit ($R~\sim~26$) is
consistent with previous ground-based observations. At fainter
magnitudes the clustering amplitude continues to decrease but at a
slower rate than that predicted by the power law $w(1'')\propto
10^{-0.27R}$ observed for shallower samples.  The observed \w over the
magnitude range $20\ <R<\ 29$ is consistent with linear evolution of
the clustering of a galaxy population which at present has a
correlation length $r_0$ of  about 4 \h1 Mpc, close to that of local {\it
IRAS} galaxies.  We also investigate the impact that magnification bias
induced by weak gravitational lensing may have on our results.
Although the observed amplitude of \w can differ from the true
amplitude by up to 30\%, this effect is not large enough to affect our
conclusions. Finally, by using a color-selected sample, we examine
whether the expected effects of magnification bias can be used for an
independent determination of cosmological parameters in deep images.
We conclude that  the amplitude of the effect can be large and in some
cases even produce an upturn of the amplitude of the correlation with
limiting magnitude. However, we find that it
is  not possible to detect the effects of magnification
bias on \w from images alone.  If redshift information
becomes available,  it is possible to 
measure the effects of magnification bias directly and thus
constrain the density parameter $\Omega_0$
and the bias factor $b$.

\end{abstract}

\keywords{cosmology: observations, gravitational lensing, large-scale structure }

\section {Introduction}

An important constraint on the formation and evolution of structures
in the Universe is the three dimensional two-point correlation
function $\xi(r)$ as a function of redshift $z$.  Unfortunately,
redshift surveys to measure the intrinsic clustering properties of
faint galaxies ($R$ \gsim $25.5$), presumably at high redshifts ($z
\gsim 1$), are 
difficult even with the new generation of large aperture
telescopes. Such direct measurements of $\xi(r)$ for $z \sim 1$ are
only now becoming available (\eg Cole
\etal 1994, Le F\`evre \etal 1996). Therefore, to study the clustering
properties of faint galaxies, one must for now  rely on studies of the 
angular two-point correlation function \w. Some constraints on 
the redshift dependence of $\xi(r)$ can be
obtained by investigating the dependence of the amplitude of \w
on the limiting magnitude.

Recent studies of \w have pushed the limiting magnitude to ever
fainter flux levels. Current observational limits reach $R =26$ (\eg
Brainerd, Smail \& Mould 1995, hereafter BSM). There is a general
agreement that the amplitude of \w decreases more rapidly with
limiting magnitude than
expected from a redshift distribution $N(z)$ as predicted by ``no-evolution models'' and
linear evolution of the clustering.  However, the interpretation of these
observations depends on the assumed model for the redshift distribution
and clustering evolution, which are both poorly constrained by current
data (\eg Glazebrook \etal 1995, Le F\`evre \etal 1996). Some authors
have argued that good agreement with the data can be obtained with
models which assume modest clustering evolution of locally observed low
surface brightness galaxies (BSM).

Extending the analysis to fainter magnitudes is of great interest in
order to impose more stringent constraints on the epoch of formation
of structures. However, as first pointed out by Villumsen (1996,
hereafter V96), the interpretation of such data needs to take into
account the effect of magnification bias induced by weak gravitational
lensing, which may affect the measurement of \w at faint
magnitudes. As discussed by V96, this effect is expected to be
important for samples with a median redshift $\gsim$ 1 and it may
therefore affect the analysis of very deep galaxy samples such as
those extracted from the Hubble Deep Field (hereafter HDF, Williams
\etal 1996). The effect should be most evident in samples which
preferentially include ``red'' galaxies and therefore have shallow
number counts slopes (Broadhurst 1996).  A possible signature of the
effect would be an upturn of the correlation amplitude with the median
redshift of the sample which should correlate with the magnitude
limit.

The detection of magnification bias could be an important tool to
further constrain cosmological models.  The amplitude of this effect
is a measurement of the clustering of the mass and depends on the
product $\Omega_0 \sigma_8$, where $\Omega_0$ is the cosmological
density parameter and $\sigma_8^2$ is the variance of the {\it mass}
fluctuations within a sphere 8 \h1 Mpc in radius. Therefore, the
behavior of the clustering amplitude as a function of the limiting
magnitude could provide a test on the value of the product $\Omega_0
\sigma_8$.

In this paper we use the HDF to investigate the behavior of the
amplitude of \w with magnitude reaching at least 3 magnitudes fainter
than published data from ground-based observations. The faint
magnitude limit and the color information make these data ideal,
except for the small angular coverage, to investigate the behavior of
\w at faint flux levels and the contribution of magnification bias to
the observed clustering of faint galaxies.  Previous work on galaxy
clustering in the HDF field has been carried out by Colley \etal
(1996) focusing on very small angular scales and discussing the
possible existence of sub-galactic clumps at high redshift.  Here,
instead, we use the HDF data to investigate the evolution of $\xi(r)$
as a function of redshift.

In section 2 we describe the  catalog used in the analysis.  In
section 3 we predict the correlation function \w in the absence of
magnification bias for simple models of the redshift distribution and
clustering evolution. In section 4, we compute the two-point angular
correlation function for different magnitude-limited samples and
determine the variation of the clustering amplitude as a function of
the magnitude limit.  Section 5 describes the theoretical calculation
of the effects of magnification bias on \w, and compares the modified
curves with the data. In section 6, the same analysis is done for a
color-selected sample which should be more sensitive to the effects of
the magnification bias.  Our conclusions are summarized in section 7.

\section{The Galaxy Catalog}

Williams \etal (1996) presented a catalog of galaxies extracted from
the HDF images with FOCAS. An alternative galaxy catalog has been
used by Clements \& Couch (1996). 
This catalog which was generated using
the SExtractor program (Bertin \& Arnouts, 1995),
was kindly provided to us by Couch (1996). As
pointed out by Williams \etal, the FOCAS catalog 
finds in a significant number of cases several objects where visual
inspection of the images indicates that there is only a single galaxy.
A comparison of the two catalogs with the HDF images leads us to believe
that Couch's catalog contains a smaller number of
such cases.  For this reason, we used that catalog for this work.  A
minimum object extraction area of 30 pixels and a detection threshold
of 1.3 $\sigma$ above the background were used.  The magnitudes were
computed with the zero points given by Holtzman \etal (1996). For this
work, we used the catalogs extracted from the images taken with the
F606W filter, which is similar to an R passband, and the F814W filter,
which corresponds to the I band.  Since only a very small number of
stars is expected in the HDF field, we did not attempt to use the
galaxy/star separation parameter given by the SExtractor program and
treated all detected objects as galaxies.  In order to avoid edge
problems, only galaxies with pixel coordinates in range $250 < x,y <
2050$ were used.  The total of 1732 galaxies detected in the F606W
filter were used, out of which 1256 were detected in both the F606W and
F814W filters. Hereafter, we refer to the two bands as R and I,
respectively.

\section{Predicted Correlation Function}

The method to derive the angular correlation function from the
intrinsic correlation function (neglecting magnification bias) for a
given redshift distribution is well known, (\eg Peebles 1980).
Following Efstathiou \etal (1991) we assume that the evolution of the
intrinsic correlation function is given by

\be
\xi(r,z)=\left({r \over r_0}\right)^{-\gamma}(1+z)^{-(3+\epsilon)}\ ;\
\gamma=1.8\ , \label{xi}
\ee
when expressed in proper coordinates. A power index $\epsilon =0.8$
corresponds to linear evolution of the correlation function, while
$\epsilon=0$ corresponds to a correlation function constant in proper
coordinates.  Here $r_0$ is the present day correlation length.

We adopt the redshift distribution \be N(z)={\beta z^2 \over z_0^3
\Gamma[3/\beta]} \exp\left[-\left(z/z_0\right)^\beta\right] \ ;\
\beta=2.5\ , \label{nz} \ee where $z_0$ is approximately the median
redshift (\eg Efstathiou \etal 1991) and $\Gamma$ is the Gamma
function.  Values for $z_0$ were provided by Charlot (1996).
The adopted redshift distribution predicts that for a
magnitude limit of $R=28$, 84\% of the galaxies are at $z > 1$ and 29\%
of the galaxies at $z > 2$. Given the uncertainties, these
estimates are in reasonable agreement with the estimated redshift
distribution based on photometric redshifts from HDF (Mobasher \etal
1996).
It is important to note that the redshift distribution and hence the
median redshifts are quite uncertain for faint, magnitude-limited
samples such as the one considered here.  This is probably the largest
uncertainty in inferring the amplitude and evolution of  $\xi(r,z)$ from
the present sample.

With this parameterization, we can calculate \w using Limber's
equation (\eg Efstathiou \etal 1991)
\bea
\omega(\theta)&=&\sqrt{\pi}\;{\Gamma[(\gamma-1)/2] \over
\Gamma[\gamma/2]}\; r_0^\gamma\; \theta^{1-\gamma}
\int_0^\infty dz \nonumber \\ 
&&H(z)\;N^2(z)\times x^{1-\gamma}\;(1+z)^{\gamma-3-\epsilon}. \label{ampl1}
\eea
Here $H(z)$ is the Hubble constant as a function of redshift,
normalized so that $H(z=0)\equiv 1$, and
$x(z)=2\left(1-(1+z)^{-\half}\right)$ is the comoving angular diameter
distance.  Since the results for the observed correlation
function depend only weakly on $\Omega_0$ (BSM), we assume 
hereafter that $\Omega_0=1$.

\section{Correlation Function from the HDF Catalog}

We have extracted from our catalogue eight R magnitude-limited samples
with  magnitude limits ranging from $R=26.0$ to $29.5$ in 0.5
magnitude steps, discarding all galaxies brighter than $R=23$.  
Although the samples as defined are not totally
independent, they are nearly so because the sample size increases rapidly with
limiting magnitude. 

The angular correlation function \w is estimated using the estimator 
described by Landy and Szalay (1993) and BSM
\be
w (\theta)={DD -2DR+RR \over RR}. \label{DD}
\ee
Here $DD$, $DR$, $RR$ are the number of data-data, data-random, and
random-random pairs at a given angular separation.  

For each magnitude limit we generated a random sample within the same
region of the galaxy catalog but with five times as many objects.  The
number of $DR$ and $RR$ pairs are scaled to the number of $DD$ pairs.
The angular correlation function \w for each magnitude limit was
estimated from pairs of galaxies with angular separations in the range
$2^{``} < \theta < 80^{``}$. The total number of pairs at a given
separation was calculated by summing the number of pairs at the
corresponding separation within each individual chip.  
Pairs across chip boundaries were not included  in order to minimize
additional uncertainties associated with 
chip-to-chip variations of the photometric zero-point.
The finite number of galaxies in the random samples add an uncertainty
to the observed correlation function.  However, this uncertainty is
far smaller than the uncertainty due to the finite number of real
galaxies.  A recalculation of \w with ten times as many random
galaxies as real galaxies lead to the same result.
As a consistency check, we have also estimated \w from counts in
cells and the results are similar to those shown here.

Errors were estimated from both Poisson statistics and from 30
bootstrap resamples of the data (see \eg Barrow, Sonoda \& Bhavsar
1984).  Both error estimates agree well at large separations.  Fits to
the correlation function are not sensitive to the differences between
the error estimates at small scales.  The errors estimated from the
bootstrap resampling are consistent with the field-to-field variations
of \w as measured for the three chips.

Due to the small angular size of the chips, it is necessary to 
take into account
the ``integral constraint''.  The background density of galaxies is
estimated from the sample itself, forcing the integral of the
correlation function over the survey area to be zero. Since the
angular size of the survey area is small, \w is reduced by the amount
\be
C\equiv {1 \over \Omega^2}\int\int d\Omega_1 d\Omega_2 w(\theta ), \label{iconst}
\ee

\noindent the so-called integral constraint (BSM).
  Here $\Omega$ is the solid angle of the survey
area on the sky.
If we assume that \w is a power law
\be
w(\theta )=A\; \theta^{-\gamma+1}\ ;\ \gamma=1.8, \label{wtheta}
\ee
then $C=0.071 A$ for our survey geometry for $\theta$ measured in
arcseconds.

Figure 1 shows the observed \w for the eight magnitude limits. We determined
the amplitude of \w by fitting
\be
w(\theta )=A\; \theta^{-\gamma+1}-C\ ;\ \gamma=1.8, \label{wthetaobs}
\ee
which takes into account the integral constraint. These fits are shown
as solid lines in figure~1.  The data points at different angular
separations are correlated but this is ignored in the fit.  The error
bars represent 1-$\sigma$ Poisson errors.  There are two possible
systematic effects that may affect our \w estimates.  The first is due
to merging of images, especially at fainter magnitudes.  This effect
should not be important in our calculations as they will
preferentially affect \w on very small angular scales which have
little weight on our fits to the data.
% On larger angular scales we
%expect this effect to be gradually less important.
%e considerably smaller and would decrease the
%observed \w below the real \w.  Thus our results are, if anything, an
%underestimate of the real \w.
A second potential effect is due to the incompleteness of the galaxy
sample close to the magnitude limit of the HDF data.  Although there
is no detailed study of the incompleteness of the HDF galaxy sample,
the behavior of the number counts down to $R \sim 29$ indicates that
incompleteness should not be important at brighter magnitudes.  Moreover,
the conclusions presented below would not change if we considered only
galaxies brighter than $R\sim 28$ which is 1 magnitude brighter than
the magnitude at which we may reasonably expect the onset of
incompleteness effects.

We have applied Kendall's
$\tau$ test to estimate the significance
level at which the null hypothesis of zero signal can be rejected.  We
find that the significance of the detection is 
$>2\sigma$ for all subsamples, except the brightest, presumably
because the small number of galaxies, and the faintest, possibly
because of the incompleteness of the sample.  In Table 1 we summarize
the observational results.  The first five columns give the magnitude
limits, the estimate of the median redshift, the number of
galaxies in the sample, the amplitude of \w at 1'', and the associated
$1\sigma$ error.

Figure 2 shows the amplitudes of the correlation function as derived
from the fits of figure~1 as a
function of magnitude limit, scaled to a separation of 1~arcsec.
The filled circles are the HDF data and the
error bars are 1-$\sigma$ uncertainties from fitting
Eq.~(\ref{wthetaobs}) to the measured correlation
function.  The open circles
are the data taken from BSM but converted to the amplitude at 1
arcsec.  The figure includes data from Couch \etal (1993),
Efstathiou \etal (1991), Roche \etal (1993), and Stevenson \etal
(1985), in addition to the BSM data points.

The results of the HDF data are consistent with those of BSM at bright
magnitudes, despite the admittedly large error bars. However, at
fainter limits the clustering amplitude falls off slower than the
power-law $A(R)\ \propto\ 10^{-0.27\;R}$ proposed by BSM to fit 
the data at brighter magnitudes in the range $18\ <R<\ 25$.  Although
all HDF points lie well above this line there is no indication of an
upturn at very faint magnitudes as previously claimed by different
authors in the B-band (\eg Neuschaefer \etal 1991, Landy, Szalay, \& Koo
1996).

The curves in figure 2 are theoretical predictions of the amplitude of
\w assuming a redshift distribution given by equation \ref{nz}, using
values for the median redshift as listed in Table I. We have
considered models with $\epsilon=0$ and $0.8$, and $r_0=2$ to
$5\;h^{-1}$ Mpc in $1\;h^{-1}$ Mpc steps.  The dotted curves are the
predictions for a local galaxy population with different correlation
lengths $r_0$ and $\epsilon=0$, while the solid curves are the
predictions for $\epsilon=0.8\,$.  The theoretical curves predict that
the correlation amplitude will decrease slower than the
power law observed at brighter magnitudes, consistent with our data.

A model with linear evolution of the clustering amplitude
($\epsilon=0.8$) with a present day correlation length $r_0 \approx$ 4
\h1 Mpc fits the data remarkably well over a range of nine magnitudes,
$20 \lsim R \lsim 29$. This value of $r_0$ is very similar to that
measured for IRAS galaxies (Fisher \etal 1994), consistent with the
picture that faint HDF and present day IRAS galaxies are drawn from
the same population of field galaxies.  If this is true, it would
suggest that IRAS galaxies were formed at redshifts $z\ \gsim\
1.5$. As pointed out above, these conclusions depend critically on the
assumed redshift distribution and must await supporting evidence.

This conclusion contrasts to that of BSM who argue in favor of a very
weakly clustered population with $r_0 \approx$ 2 \h1 Mpc, presumably
low-surface brightness galaxies, or the
interpretation Landy, Szalay, \& Koo (1996) for a low redshift
population of very faint, weakly clustered blue galaxies.  The reason
for this discrepancy is the higher assumed median redshift, which is
supported by the results of Mobasher \etal (1996).  At faint
magnitudes, the HDF data can also be fitted with a model with slow
evolution \ie $\epsilon=0$, if $r_0=2$ to $3$ \h1 Mpc.  However, such
a model would be inconsistent with the data for $R \lsim$ 25, being
too shallow at brighter magnitudes.

Since we have assumed that the median redshift of galaxies in HDF is
larger than unity, the effects of weak gravitational lensing can
influence the observed correlation function of galaxies.  This so
called magnification bias could in principle affect our conclusion
(V96).  The importance of this effect for the HDF sample is evaluated
in the next section.

\section{Effects of Magnification Bias}

V96 has shown that the magnification bias can have a significant
influence on \w provided that: 1) the sample depth in redshift is $z
\gsim 1$; 2) the slope of the number counts $s$ is significantly
different from 0.4; and 3) the quantity $\Omega_0\sigma_8$ is not much
less than unity.  We emphasize that $\sigma_8$ is the rms mass density
fluctuations on a scale of 8 \h1 Mpc.

The magnification bias, as shown by Broadhurst, Taylor \& Peacock
(1995), will affect the number density of galaxies at a given position
on the sky.  If the slope of the number counts is $s$ and
the magnification due to gravitational lensing from a matter
overdensity is $\mu$, then the observed number density $N_{obs}$ of background
objects will be different from the ``true'' number density $N_{true}$
of objects
\be
N_{obs}=N_{true}\ \mu^{2.5s-1}. \label{broad}
\ee
Note that $\mu=1$ corresponds to no magnification.  Assuming there are
no intrinsic correlation of galaxies, then apart from random
fluctuations due to the finite number of galaxies, the number density of
galaxies is constant across the sky. In the presence of mass density
fluctuations,  gravitational lensing will change the number density of
galaxies according to Eq.~(\ref{broad}). The amplification will be a
function of position in the sky, and therefore the number density of
galaxies will be a function of position.  This means that
a non-zero correlation amplitude is observed even in the absence of
intrinsic galaxy clustering.  In the limit of weak clustering, \ie
$|\mu-1| \ll 1$, the magnification $\mu$ relates to the ``convergence''
$\kappa$ as $\mu=1+2\kappa$ (\eg Blandford \etal 1991, Broadhurst,
Taylor \& Peacock 1995).  The quantity
$\kappa$ is a dimensionless measure of the surface mass overdensity.
If the intrinsic relative overdensity is $\Delta\mu$ we find that

\be
N_{obs}=N_0 \left(1+\Delta\mu+(5s-2)\kappa\right)\ . \label{jvv}
\ee
Here $N_0$ is the average number density of galaxies so that
$N_{true}=N_0 (1+\Delta\mu)$ and $N_{obs}$ is the observed number
density of galaxies.
Since we assume that the clustering and the lensing are both weak, the
clustering term $\Delta\mu$ and the lensing term $(5s-2)\kappa$ are additive
and we get an observed correlation function
\bea
w(\theta
)&\equiv&\left<{N_{obs}(\vec\theta_0)N_{obs}(\vec{\theta_0}+\vec\theta)
\over N_0^2}-1\right> \nonumber \\
&\equiv&\omega_{gg}(\theta)+
(5s-2)^2\omega_{\kappa\kappa}(\theta)+\nonumber\\
2(5s-2)\omega_{g\kappa}(\theta).
\label{wthetajvv} 
\eea
The brackets denote a directional average.
It can be seen that there are three terms to the observed correlation function.
The first term is due to the true clustering of galaxies.  The second
term is due to the mass density fluctuations while the last is a cross
term.  For details of the calculation of the effects of gravitational
lensing, see V96.

In order to evaluate the importance of magnification bias, 
we compare the two terms in Eq. \ref{jvv} on
a scale of 5 arcmin. 
The rms density fluctuation on that scale is
approximately 0.035 for the sample $R < 25.5$ (BSM).  If the redshift
of that sample is $z \sim 1$ then the rms $\kappa$ smoothed on the same
scale is $\sigma_\kappa \approx 0.015 \Omega_0 \sigma_8$.  If $s=0.3$
then the lensing term in Eq. \ref{jvv} will be approximate
$(5s-2)\kappa \approx 0.008$ for $\sigma_8\Omega_0 =1$.  This
is not negligible compared to the intrinsic clustering term
unless $\Omega_0\sigma_8\ll 1$.  If our galaxy sample has
a number-count slope of $s=0.2$, we greatly increase the effect of
lensing.  In that case, the lensing term $(5s-2)\kappa$ will double to
$\approx 0.016$.  Therefore, \w can be significantly influenced
by magnification bias as shown in fig.~1 of V96. 
In that paper the effect was evaluated in terms of median redshift of
the sample.

For a given set of cosmological parameters, such as $\Omega_0$,
$\Lambda$, $\sigma_8$, and intrinsic galaxy-galaxy $\xi_{gg}(r,z)$ and
mass-mass $\xi_{mm}(r,z)$ correlation functions we can calculate
$\omega_{gg}(\theta)$ and $(5s-2)^2\omega_{\kappa\kappa}(\theta)$ for
an observed redshift distribution $N(z)$ of galaxies.  In order
to calculate $2(5s-2)\omega_{g\kappa}(\theta)$,
it is necessary to
compute the 
galaxy-mass $\xi_{gm}(r,z)$ correlation function, or in
other words, the relation between the galaxy and mass distributions.
If there is no
correlation, then $\omega_{g\kappa}$ is by definition zero.  The next
simplest assumption is the linear bias model, whereby the galaxy
overdensity equals the mass overdensity apart from a factor $b$.  This
is the model we use for theoretical predictions,
\be
\xi_{gm}(r,z)=b\ \xi_{mm}(r,z)\; ; \xi_{gg}(r,z)=b^2\;\xi_{mm}(r,z),
\label{corrs3} 
\ee
where $b$ is the biasing factor. 
In the linear bias model we then get (V96)
\bea
&&\omega(\theta)=\sqrt{\pi}\;{\Gamma[(\gamma-1)/2] \over
\Gamma[\gamma/2]}\; r_0^\gamma\; \theta^{1-\gamma}
\int_0^\infty dz \Bigl(\;H(z)\;\nonumber\\
&&\left[\;N(z)+\;3\;{\Omega_0 \over b}\;[5s-2]\;\phi(z)\;y(z)\;
[1+z]\;H^{-1}(z)\right]^2 \nonumber \\
&\times&x(z)^{1-\gamma}\;(1+z)^{\gamma-3-\epsilon} \Bigr). \label{wmb}
\eea
This is a generalization of equation \ref{ampl1}.  Here $x(z)$ and $y(z)$ are
the comoving radial and angular diameter distances and $\phi(z)$ is
the lensing selection function which is the integral over all sources
more distant than $z$ of the ratio of the comoving angular lens-source
distance $y_{LS}(z',z)$ and observer-source distance $y_{OS}(z)$
\be
\phi(z')=\int_{z'}^{\infty}dz\; N(z){y_{LS}(z',z) \over y_{OS}(z)}. \label{phi}
\ee

As stated in V96, the cosmological constant $\Lambda$ has little
influence on weak gravitational lensing and therefore magnification
bias.  In Eq. \ref{wmb} the dependence of \w on the rms mass density
fluctuations parameterized by $\sigma_8$ has been expressed in terms of
the galaxy correlation length $r_0$ and the bias factor $b$ for a
particular galaxy population, using the linear bias model.

The contribution to \w from gravitational lensing depends 
on the product $\Omega_0 \times \sigma_8$.  For the cross term, the
dependence is linear and for the pure lensing term the dependence is
quadratic.  In this paper 
we have performed all calculations for a model in which $\Omega_0=1\
;\ \Lambda=0$ and for various values of
$\sigma_8$.  We note that for a fixed value of $r_0$, a scaling in
$\sigma_8$ is equivalent to a scaling in $b$.

In Figure 3 we show the theoretical predictions for the correlation
amplitude including the effects of magnification bias.  The slope of
the number counts in the $R$ band is measured to be $s=0.31 \pm 0.03$,
consistent with results from brighter samples (Smail \etal 1995).
The observed data points are shown in the two panels of Figure 3
together with the theoretical predictions both with and without
magnification bias.  In the left panel we assume that $\epsilon=0$ and
show the predictions for the amplitude for $\sigma_8=0.5$, \ie a
highly biased model (long-dashed
curve) and for $\sigma_8=1$ (short-dashed curve), \ie a unbiased
model.  In the right panel
it is assumed that $\epsilon=0.8$.

These theoretical curves can be compared with the curves in V96,
Figure 1.  In that paper, the effects of magnification bias are shown
in terms of median redshift, while here we show the effects in terms of
limiting magnitude which is an observable.

The basic effect of magnification bias is to decrease the observed amplitude of
clustering.  It is larger for small values of $r_0$, and larger
values of $\epsilon$.  The effects of magnification bias is also an
increasing function of limiting magnitude.  These effects can be simply
understood in terms of a competition between the effects of true
clustering and the clustering induced by the magnification bias.  The
intrinsic clustering contribution decreases with decreasing value of
$r_0$ and increasing values of $\epsilon$ and limiting magnitude.  The
clustering contribution from magnification bias is an increasing
function of $\sigma_8$ and limiting magnitude.  From the figure, we find
that the decrease in amplitude is in the range of 10\% to 25\% and is
not very sensitive to the limiting magnitude.  The amplitude is already
high at a magnitude limit $R\ \sim\ 25$ which corresponds to the
spectroscopic limit of large telescopes.  As discussed below, this fact
provides us with a possible way of directly measuring the magnification
bias and therefore $\Omega_0 \sigma_8$ in deep redshift surveys.  
Even though the effects of magnification bias are not negligible, 
they are not large enough to change the conclusions in section~4.

In our analysis we have considered only the effect of magnification
bias caused by large scale structure.
Weak lensing by individual galaxy halos, \ie galaxy-galaxy lensing
(\eg Brainerd, Blandford, Smail 1996) can change the correlation
function on small angular scales through the magnification bias.
Given the uncertainty in
the size and masses of galaxy halos, the magnitude of this effect is
uncertain.  However, judging by the results of Brainerd, Blandford, \&
Smail (1996), which reported a weak detection of the effect, it is
unlikely that this effect will make a significant contribution to \w.

The reason the effects of magnification bias are not more noticeable
is that there is a large fraction of low redshift galaxies in the
sample for which magnification bias is not important.  Furthermore,
the full sample considered so far has a number-count slope $s\approx 0.3$,
close to the neutral slope $s=0.4$.  This motivates us to look for a
subsample for which the number-count slope is significantly lower than 0.3.

\section{Magnification Bias for a Color-selected Sample}

It is clear from Eq. \ref{wthetajvv} that the effects of magnification
bias are larger if the number counts slope for the sample considered,
$s$, is significantly different from 0.4~.  In particular V96 has
shown that for a sample with $s=0.2$, magnification bias can produce a
distinctive upturn in the amplitude of \w for samples with median
redshifts $z > 1$.  If the median redshift increases with limiting
magnitude of the sample, then we can in principle expect to see this
distinctive upturn at faint magnitudes.  

It has been noted by Broadhurst (1996) that the reddest galaxies, as
defined by $V-I$ colors, have a small number-count slope of $s < 0.2$.
Moreover Broadhurst \etal (1996) have shown that the number counts
slope is a decreasing function of $V-I$ color, \ie red-selected
samples have a shallower slope.  Bearing this in mind, we define color
selected subsamples such that the slope of the number counts is
$0.2\,$.  This was done as follows. For a given magnitude limit in
$R$ we arranged the galaxies in order of increasing $R-I$ color index.
We removed blue galaxies until the slope of the number counts of the
remaining sample is $s=0.2\,$  and calculated \w in the same way as
for the full sample.

The results are shown in Figure 4 in the same form as in Figure 1, and
are listed in columns 6 to 8 of Table 1.  As for the full sample, we
detect a correlation signal down to $R\ \sim\ 29$.  Figure 5 is
equivalent to figure 3 for the full sample except that there are no
BSM data points as they have not color-selected  samples.  The
dotted curves are the same for the full sample shown in Figure~3, as
they do not include the magnification bias. We have assumed the same
redshift distribution for the color selected sample for illustrative
purposes only.  The two samples need not have the same redshift
distribution.  The dashed curves include the effects of magnification
bias.

The theoretical predictions for the correlation amplitude show the
expected larger influence of the magnification bias.  The major
differences in including the magnification bias for the red-selected
sample are: 1) At brighter magnitudes, $R\ \lsim\ 25$, the effect is
significantly larger than for the full sample, but has a similar
behavior.  2) At faint 
%HHHHHHHHHHHHHHHHHHHHHHHHH
magnitudes the effects are much more complicated.  For $\sigma_8=0.5$
models, the observed effect of magnification bias for the color
selected sample is larger than for the full sample, but again not enough to
significantly affect the conclusions about $r_0$ for a given
%HHHHHHHHHHHHHHHHHHHHHHHHH
$\epsilon$.  For $\sigma_8=1$ models, the behavior is quite different with
the curves flattening out and in some cases showing an upturn at faint
magnitudes.  Unfortunately, the upturn is not a generic feature but
occurs only for models with weak intrinsic clustering.
Furthermore, an upturn in the correlation amplitude at faint
magnitudes could have other explanations, such as a very faint local
population of galaxies.

It is important to note that the effects of magnification bias can be
larger at brighter magnitudes than at fainter magnitudes.  This is
important as it allows us to use spectroscopic information to 
measure the magnification bias.  There are several ways of
implementing this.  If we have some redshift information, either
photometric or spectroscopic, physical pairs can be removed from the
sample and the observed $w(\theta)$ would be a direct measure of the
magnification bias.  An alternative way is having spectroscopic
redshifts for a random subsample so that we can derive $N(z)$ and
$\xi(r,z)$ directly.  From this information we can compute the true
$w(\theta)$ from Limber's equation and compare this result with the
measured $w(\theta)$.  The difference would then be the contribution
from magnification bias from which we can measure $\Omega_0 \sigma_8$.

\section{Conclusions}

Analysis of the multicolor HDF data using the two-point angular
correlation function leads us to the following conclusions:

\ni 1- We detect a clustering signal down to a magnitude limit of 
$R=29$. 

\ni 2- We find  values of the amplitude of the \w consistent with
those obtained from ground-based observations at $R \sim 26$.
The amplitude continues to decrease down to the faintest
magnitude limit considered with a slope  consistent with $\gamma \sim
1.8$.

\ni 3- Our results show that the measured amplitudes of \w are 
consistent with some of the theoretical models.  The best fit model is
one with linear evolution of the correlation function \ie $\epsilon
=0.8$ and a present day correlation length $r_0 \sim 4$ \h1 Mpc
similar to that observed for IRAS galaxies.  This would be consistent
with a large population of normal galaxies already being in place at $z\lsim
1.7\,$.

\ni 4- We also show that even though magnification bias is expected to have
an impact on the angular correlation function at faint magnitudes, it is not
strong enough to seriously affect the the conclusions about $r_0$ and
$\epsilon$, even for a red selected sample, for which magnification
bias should be more important.

\ni 5- The effects of magnification bias can be important for samples
with a limiting magnitude of $R\ \lsim\ 25$, which corresponds to the
spectroscopic limit of large telescopes. This makes it possible to
measure the effects of magnification bias and therefore get an
alternative way of measuring $\Omega_0 \sigma_8$.
 Given the estimates for the magnification bias, this requires a galaxy
sample large enough to allow measurements of $w(\theta)$ to better than
10\% accuracy.

Deep  imaging and redshift surveys from ground based telescopes 
will make it possible to draw more definite
conclusions about the effects of magnification bias and its potential 
use for the measurement of $\Omega$.

\bigskip

\noindent {\bf Acknowledgments}

We would like to thank  Warrick Couch for his HDF catalog of galaxies, 
and the ESO/ECF HDF group for many useful discussions.

\vfill\eject
\begin{table*}
\caption{Observed correlation functions}
\begin{tabular}{ccccccccc}
 Mag Limit&Redshift&  \# of Galaxies & Amplitude &    Error  & \phantom{xxx} &\# of Galaxies &
Amplitude &   Error  \cr 
          &Estimate& \multicolumn{3}{c}{F u l l\ \ \  S a m p l e}  &       &\multicolumn{3}{c}{R e d\ \ \  S a m p l e}         \cr
\hline \cr
      20.0&0.25\cr
      21.0&0.35\cr
      22.0&0.48\cr
      23.0&0.64\cr
      24.0&0.87\cr
      25.0&1.11\cr
      26.0&1.35&       263&0.19&      0.18&    &   249&	0.35&      0.19\cr
      26.5&   &       384&0.31&      0.14&    &  362&	0.30&      0.15\cr
      27.0&1.54&       536&0.06&      0.10&    &   510&	0.10&      0.10\cr
      27.5&   &       692&0.15&      0.07&    &   655&	0.08&      0.08\cr
      28.0&1.71&       900&0.13&      0.06&    &   839&	0.13&      0.06\cr
      28.5&   &      1230&0.071&     0.043&   &   1082&	0.073&      0.049\cr
      29.0&1.87&      1559&0.063&     0.033&   &   1226&	0.057&      0.043\cr
      29.5&   &      1732&0.059&     0.030&   &   1256&	0.043&      0.041\cr
\hline\cr
\end{tabular}
\end{table*}

\newpage

\noindent {\bf Figure Captions}
\bigskip

{\bf Figure 1.} Measured $w(\theta)$ for all galaxies brighter than
the limiting magnitude R as indicated in each panel.  Error bars
represent the $1\sigma$ Poisson errors. The solid lines are the best
fits for fixed power law index $\gamma=1.8$. The integral constraint
(see text) has been included in the fits.

{\bf Figure 2.}  The observed amplitude of $w(\theta)$, measured at 1
arcsec separation as a function of limiting R magnitude, for $18\ < R
<\ 29$.  Filled circles are from HDF data and open circles are adapted
from Figure~2 of BSM, including 1~$\sigma$ error bars are shown.  The
dashed line is a power-law fit to the data from BSM.  The curves are
theoretical predictions assuming $N(z)$ as described in the text for
different values of the present correlation lengths $r_0$ as
indicated, with the amplitude increasing with the adopted $r_0$.  The
dotted curves are for $\epsilon=0$, while the solid curves are for
$\epsilon=0.8\,$.

{\bf Figure 3.} Observed amplitude of $w(\theta)$, measured at 1
arcsec separation as a function of limiting R magnitude, for $24\ < R
<\ 29$.  Points with filled circles are from HDF data and open circles
are from BSM data.  The curves are theoretical predictions assuming
$N(z)$ as described in the text for different values of the present
correlation lengths $r_0$ as indicated.  In the left panel
$\epsilon=0$ is assumed while in the right panel $\epsilon=0.8$ is
assumed.  The dotted curves are predictions without the effects of
magnification bias.  The dashed curves are equivalent curves including
the effects of magnification bias.  The long dashed curves are for
$\sigma_8=0.5$ and the short dashed curves are for $\sigma_8=1$.  The
observed number counts slope $s=0.3$ is used in the predictions.

{\bf Figure 4.} Measured $w(\theta)$ for the red selected galaxies
brighter than the limiting magnitude R as indicated in each panel.
Error bars represent the $1\sigma$ Poisson errors. The solid lines are
the best fits for fixed power law index $\gamma=1.8$. The integral
constraint (see text) has been subtracted from the fits.

{\bf Figure 5.} Observed amplitude of $w(\theta)$,
measured at 1 arcsec separation as a function of limiting R magnitude,
for $24\ < R <\ 29$ for the red-selected sample. The data points are
from the HDF data sample only.    The
curves are theoretical predictions assuming $N(z)$ as described in the
text for different values of the present correlation lengths $r_0$ 
as indicated.  In the left panel $\epsilon=0$ is assumed while in the
right panel $\epsilon=0.8$ is assumed.  The dotted curves are
predictions without the effects of magnification bias.  The
dashed curves are equivalent curves including the effects of
magnification bias.  The long dashed curves are for $\sigma_8=0.5$ and
the short dashed curves are for $\sigma_8=1$.  The observed number
counts slope $s=0.2$ is used in the predictions.


\begin{thebibliography}{99}

\baselineskip 0.4cm
 

\bibitem{} Bertin, E., Arnouts, S., 1995,  A\&AS, 117, 393.

\bibitem{} Barrow, J.D., Sonoda, D., Bhavsar, S.P., 1984, MNRAS 210, 19.

\bibitem{} Blandford, R.D., Saust, A., Brainerd, T.G., Villumsen, J.V. 1991,
MNRAS, 251, 600.

\bibitem{} Brainerd, T. G., Smail, I. R.,  Mould, J. R , 1995,  MNRAS,
275, 781.(BSM)

\bibitem{} Brainerd, T.G., Blandford, R.D., Smail, I., 1996, ApJ, 466,
623.

\bibitem{} Broadhurst, T.J., Taylor, A.N., Peacock, J.A., 1995, ApJ, 438, 49.

\bibitem{} Broadhurst, T.J., 1996, ApJ, in press.


\bibitem{} Broadhurst, T.J., Villumsen, J.V., Smail, I., Charlot,S.,
1996, in preparation.

\bibitem{} Charlot, S., 1996, private communication

\bibitem{} Couch, W.J., 1996, Private Communication

\bibitem{} Clements, D.L., Couch, W.J., 1996, MNRAS 280, L43

\bibitem{} Cole, S., Ellis, R. S., Broadhurst, T. J., Colless, M. M.,
1994, MNRAS, 267, 541.

\bibitem{} Colley, W.N., Rhoads, J.E., Ostriker, J.P., Spergel, D.N.,
1996, preprint 

\bibitem{} Couch, W.J., Jurcevic, J.S., Boyle, B.J., 1993, MNRAS,
260, 241.

\bibitem{} Efstathiou, G., Bernstein, G., Katz, N., Tyson, J.A. 
Guhathakurta, P., 1991, ApJ, 380, L47.

%\bibitem{} Fall, S.M., Charlot, S., Pei, Y.C. 1996, ApJ, in press

\bibitem{} Fisher, K.B, Davis, M., Strauss, M.A., Yahil, A., Huchra,
J., 1994, MNRAS 266, 50.

\bibitem{} Glazebrook, K, Ellis, R., Colless, M., Broadhurst, T., 
Allington-Smith, J., Tanvir, N., 1995, MNRAS 273, 157.

\bibitem{} Holtzman, J.A., Burrows, C.J., Casertano, S., Hester, J.J.,
Trauger, J.T.,Watson, A.M., Worthey, G., 1996, ApJ, submitted.

\bibitem{} Landy, S. D., Szalay, A. S. , 1993, ApJ, 412, 64.

\bibitem{} Landy, S.D., Szalay, A.S., Koo, D.C. , 1996, ApJ, 460, 94

\bibitem{} Le F\`evre, O., Hudon, D., Lilly, S. J., Crampton, D., Hammer, F.,
Tresse, L. , 1996, ApJ, 461, 534

\bibitem{} Mobasher, B., Rowan-Robinson, M., Georgakakis, A., Eaton,
N., 1996, preprint.

\bibitem{} Neuschaefer, L.W., Windhorst, R.A., Dressler, A. , 1991,
ApJ, 382, 32.


\bibitem{} Peebles, P.J.E., 1980, ``The Large Scale Structure of the
Universe'', Princeton University Press.

\bibitem{} Roche, N., Shanks, T., Metcalfe, N., Fong, R., 1993,
MNRAS, 263, 360.

\bibitem{} Smail, I., Hogg, D.W., Yan, L., Cohen, J.G, 1995, ApJ, 449, L105.

\bibitem{} Stevenson, P.R., Shanks, T., Fong, R., MacGillivray, H.T.,
1985, MNRAS, 213, 953.

\bibitem{} Villumsen, J.V., 1996, MNRAS, submitted.(V96)

\bibitem{} Williams, R.E.,  Blacket, B., Dickinson, M., Dixon, V.D.,
Ferguson, H., Fruchter, A.S., Giavalisco, M., Gilliland, R.L., Heyer,
I., Katsanis, R., Levay, Z., Lucas, R.A., McElroy, D.B., Petro, L,
Postman, M., 1996, ApJ, submitted

\end{thebibliography}
\end{document}